\documentclass[aps,prl,twocolumn,superscriptaddress]{revtex4}
\usepackage{amssymb}
\usepackage{graphicx}
\usepackage{amsmath}

\begin{document}

\title{Magnetic Order Driven Topological Transition in the Haldane-Hubbard Model }
\author{Wei Zheng}
\affiliation{Institute for Advanced Study, Tsinghua University, Beijing, 100084, China}
\author{Huitao Shen}
\affiliation{Institute for Advanced Study, Tsinghua University, Beijing, 100084, China}
\author{Zhong Wang}
\email{wangzhongemail@gmail.com}
\affiliation{Institute for Advanced Study, Tsinghua University, Beijing, 100084, China}
\author{Hui Zhai}
\email{hzhai@mail.tsinghua.edu.cn}
\affiliation{Institute for Advanced Study, Tsinghua University, Beijing, 100084, China}
\date{\today}

\begin{abstract}

In this letter we study the Haldane model with on-site repulsive interactions at half-filling. We show that the mean-field Hamiltonian with magnetic order effectively modifies parameters in the Haldane Hamiltonian, such as sublattice energy difference and phase in next nearest hopping. As interaction increases, increasing of magnetic order corresponds to varying these parameters and consequently, drives topological transitions. At the mean-field level, one scenario is that the magnetic order continuously increases, and inevitably, the fermion gap closes at the topological transition point with nonzero magnetic order. Beyond the mean-field, interaction between fermions mediated by spin-wave fluctuations can further open up the gap, rendering a first-order transition. Another scenario is a first-order transition at mean-field level across which a canted magnetic order develops discontinuously, avoiding the fermion gap closing. We find that both scenarios exist in the phase diagram of the Haldane-Hubbard model. Our predication is relevant to recent experimental realization of the Haldane model in cold atom system.

\end{abstract}

\maketitle

Correlation and topology are two of central topics in modern condensed matter physics. The Hubbard model of spin-$1/2$ fermions with on-site interaction is one of the most famous model that gives rise to highly nontrivial correlation effects, such as antiferromangetic order at half-filling. This model is relevant to many strongly correlated materials such as High-$T_c$ cuprate superconductors. Recently, this model has also been simulated by using ultracold fermions in optical lattices \cite{Esslinger, Bloch,Randy}. In 1988, Haldane proposed a model of noninteracting fermions in the honeycomb lattice, which can give rise to topological band structure and quantized Hall conductance without external magnetic field \cite{Haldane}. This effect is now known as the quantum anomalous Hall effect, and has been observed experimentally in magnetically doped topological insulators \cite{Xue}. The Haldane model has also been realized recently in cold atom experiment using shaking lattice technique \cite{Shaking,shaking-theory}.

Motivated by recent cold atom realization of the Haldane model, in this letter we study the Haldane-Hubbard (HH) model of spin-$1/2$ fermions. In this model each spin component experiences the same single-particle Hamiltonian described by the Haldane model. At half-filling ($N_\uparrow=N_\downarrow=N_\text{s}/2$, $N_\text{s}$ is the number of sites), the system is a topological band insulator in the noninteracting limit. We consider only on-site repulsive interaction between two spin components as in the Hubbard model. In the strongly interacting limit, the system will enter a Mott insulator phase. Therefore, as interaction increases, we expect that two things will happen: One is the development of certain magnetic order, and the other is the transition from a topological band insulator to a topologically trivial Mott insulator. A natural question is how these two phenomena influence each other. The study of this question will shed light on the interplay between correlation and topology \cite{previous}. Previously, although there have been considerable interests focusing on the Kane-Mele-Hubbard model \cite{kane_mele}, the HH model is much less investigated \cite{kou}.

In this letter we show that, at the mean-field level the magnetic
order can drive a topological transition, either by a continuous
second-order transition with fermion gap closed at finite magnetic
order, or by a first-order transition with a jump of magnetic order
parameter. Within the mean-field calculations, we show that both
scenarios exist in the phase diagram of the HH model. We then go
beyond the mean field by considering fluctuations of magnetic order.
We show that for the former case, the spin-wave fluctuations generate
effective interaction among gapless fermions at the nominal critical
point of topological transition. This interaction, if sufficiently
strong, can open up a gap and drives the transition to first-order.

\textit {The Model.} We consider the HH model on a honeycomb lattice whose Hamiltonian is given by
\begin{align}
&\hat{H}_\text{HH}=\hat{H}_\text{H}+U\sum_{i}\hat{n}_{i,\uparrow }\hat{n}_{i,\downarrow }\\
&\hat{H}_\text{H}=-t_{1}\sum_{\left\langle ij\right\rangle ,s }\left( \hat{c}%
_{i,s }^{\dag }\hat{c}_{j,s }+\text{h.c.}\right) \nonumber\\
& - t_{2}\sum_{\left\langle \left\langle ij\right\rangle \right\rangle ,s
}\left( e^{i\phi _{ij}}\hat{c}_{i,s }^{\dag }\hat{c}_{j,s }+\text{
h.c.}\right)- M\sum_{i,s }\epsilon _{i}\hat{c}_{i,s }^{\dag }\hat{c}_{i,s } \label{Haldane}
\end{align}
where $s=\pm$ refers to spin up and down, respectively, the $t_2$-term represents next nearest hopping with a nontrivial phase $\phi_{ij}=\pm \phi$ for different sublattices, and the $M$-term adds a potential imbalance between $A$ and $B$ sub lattices, as $\epsilon_{i}=\pm 1$ for $i$ belonging to $A$ or $B$ sublattices. The $t_2$ term breaks time-reversal symmetry, and the $M$-term breaks the inversion symmetry.  Both terms open up the gap at Dirac points, and for half-filling, a phase diagram (without the interaction term) including a topological transition from trivial insulator to topologically nontrivial insulator is shown in Fig. 1(a), across which the gap at one of the Dirac point is closed. For the topologically nontrivial insulator, each spin component fills the lower band with Chern number equalling $1$ and the total Chern number $\mathcal {C}=2$.  The interaction term can be decoupled as
\begin{align}
&U\sum_{i}\hat{n}_{i,\uparrow }\hat{n}_{i,\downarrow }=\frac{1}{2}U\hat{N}-\frac{2}{3}U\sum_{i}\mathbf{S}_{i}^{2}\nonumber\\
&\approx\frac{1}{2}U\hat{N}+\sum\limits_{i}\left( -\mathbf{m}_{i}\cdot \mathbf{S}_{i}+\frac{3\mathbf{m}_{i}^{2}}{8U}\right),
\end{align}
where we have introduced an on-site magnetic order parameter $4U\langle \mathbf{\hat{S}}_{i}\rangle/3 =\mathbf{m}_{i}$. Thus the mean-field Hamiltonian $H_\text{MF}$ is given by
\begin{equation}
\hat{H}_\text{MF}=\hat{H}_\text{H}-\sum_{i}\mathbf{m}_i\cdot\mathbf{S}_i. \label{HMF}
\end{equation}

\begin{figure}[t]
\includegraphics[width=3.45 in]
{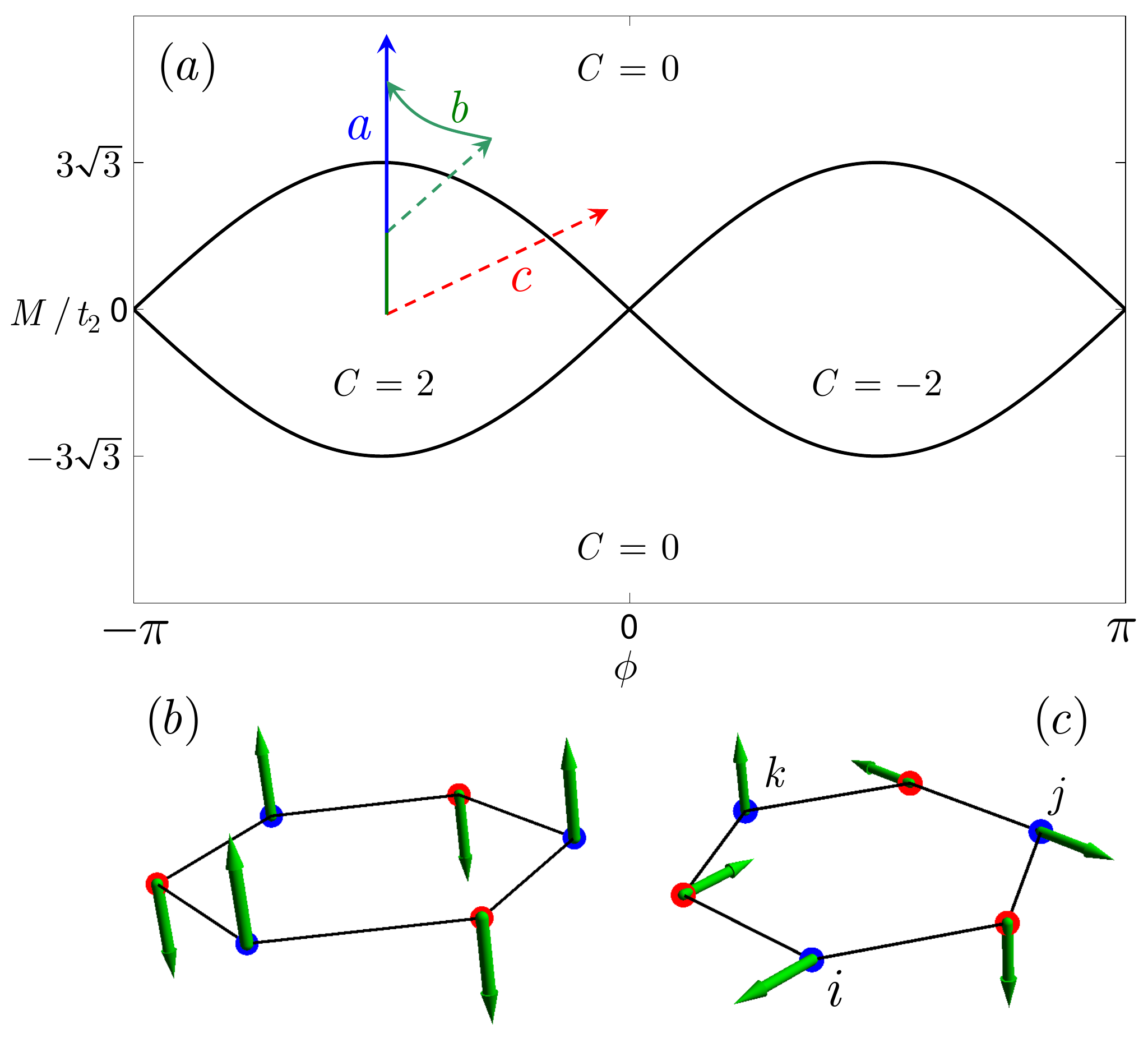}
\caption{(a) Phase diagram for the noninteracting Haldane Hamiltonian $\hat{H}_\text{H}$ (Eq. \ref{Haldane}), in terms of AB sublattice energy imbalance $M$ and phase $\phi$. The three trajectories labelled by a, b and c correspond to the evolution of mean-field Hamiltonian $\hat{H}_\text{MF}$ (Eq. \ref{HMF}) with the increasing of magnetic order, as $U/t_1$ increases, for different $t_2/t_1$ as labelled by (a-c) in Fig. \ref{PD}. The solid line means continuous evolution while the dashed line means discontinuous jump. Panels (b) and (c) illustrate two types of AF magnetic order: (b) is collinear AF order and (c) is canted AF order with nonzero spin chirality.    }
\label{spin}
\end{figure}

\textit{Relation between AF order and Topology.} Before proceeding to the self-consistent mean-field calculation, we would like to first discuss the relation between the following two types of possible AF order and the parameters in the Haldane model.

(A) Collinear AF order, i.e. ${\bf m}_i={\bf m}$ on the sublattice
$A$ and ${\bf m}_i=-{\bf m}$ on the sublattice $B$, as shown in Fig.
\ref{spin}(b). Due to the spin rotational symmetry, we can always
choose ${\bf m}=m\hat{z}$. Thus, it adds a spin-dependent
contribution on $M$ in the single-particle Haldane model of Eq.
\ref{Haldane}, i.e. $M\rightarrow M+sm$, where $s=\pm$ denotes spin.

(B) Canted AF order. For simplicity, we consider the situation that
${\bf m}$ are different among the three $A$ sites (denoted by $i$,
$j$ and $k$) of one honeycomb, as shown in Fig. 1(c), which leads to
finite ``scalar spin chirality'' order $\mathcal{S}= \langle {\bf
\hat{S}}_i \rangle \cdot ( \langle {\bf \hat{S}}_j \rangle \times \langle {\bf
\hat{S}}_k \rangle)$ ( Since we are concerned with magnetically ordered
state,  we do not use the usual definition, i.e. $\langle {\bf S}_i
\cdot ( {\bf S}_j \times {\bf S}_k )\rangle$. ). Within each unit
cell, ${\bf m}_i$ at $A$ site is approximately opposite to ${\bf
m}_i$ at $B$ site.

Then we can apply an on-site spin rotation $U_i$ so that $U_i^\dag
({\bf m}_i\cdot {\bf S}_i) U_i=\frac{1}{2}|{\bf m}_i|s_{iz}$ for $A$
sublattices and $U_i^\dag ({\bf m}_i\cdot {\bf S}_i)
U_i=-\frac{1}{2}|{\bf m}_i| s_{iz}$, where $s_{iz}$ is the Pauli matrix
associated to spin.  Qualitatively speaking, this local spin rotation introduces an
additional Berry phase factor $\pm\tilde{\phi}$ in the next nearest
hopping term for different sublattices, where $\tilde{\phi}$ is
approximately one sixth of the solid angle expanded by ${\bf m}_i,
{\bf m}_j, {\bf m}_k$, that is to say, $\phi$ in the original Haldane
Hamiltonian Eq. \ref{Haldane} should be replaced by an effective
phase $\phi_{{\rm eff}}= \phi+\tilde{\phi}$.

\begin{figure}[t]
\includegraphics[width=3.45 in]
{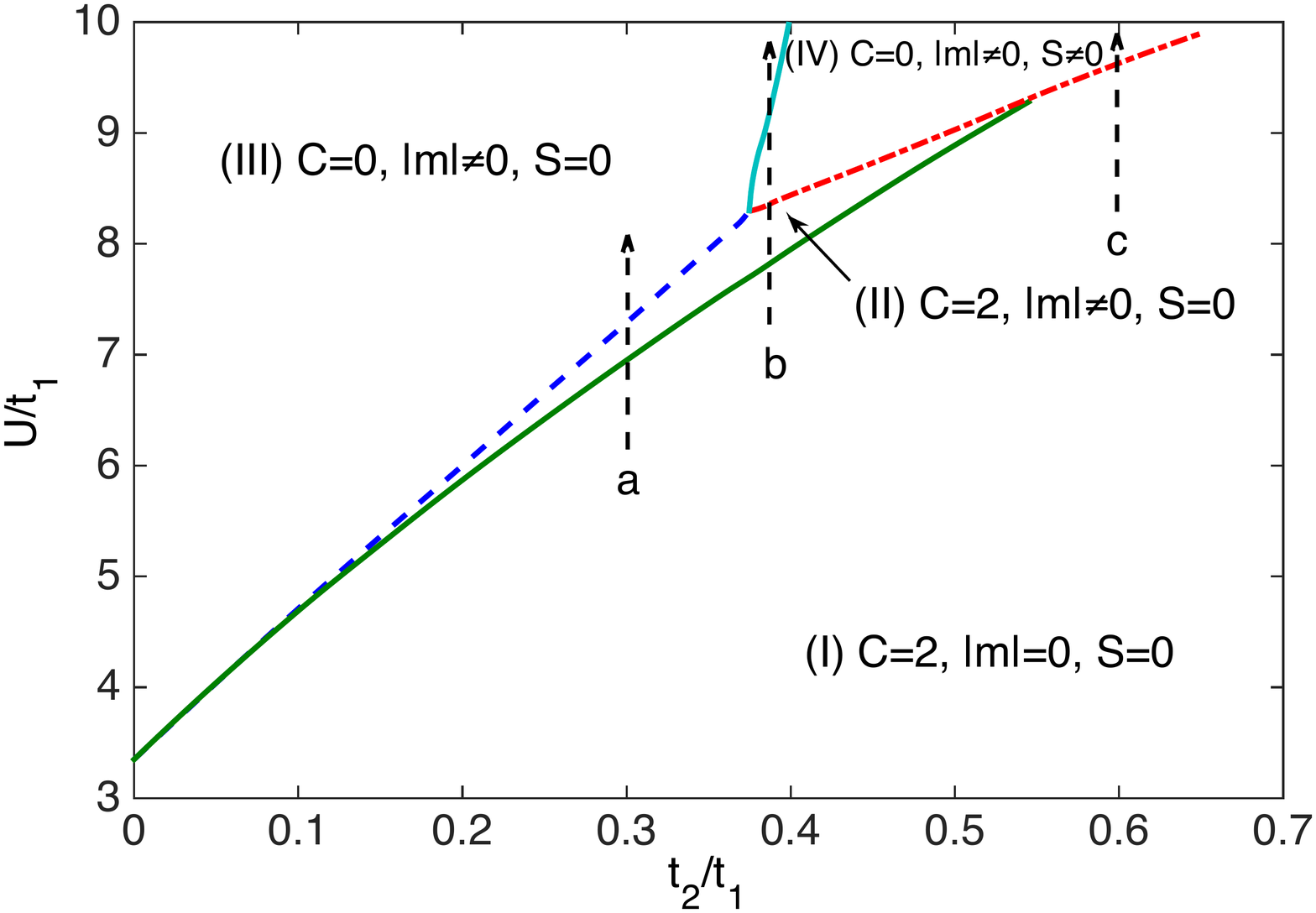}
\caption{Mean-field phase diagram for HH model in the $( t_2/t_1, U/t_1)$ plane. The green solid line is a second-order transition between a topological band insulator (Phase I) and a topological collinear AF insulator (Phase II). The blue dashed line is a topological transition between Phase II and a topological trivial collinear AF insulator (Phase III). The red dash-dot line is a first-order transition between either Phase I or Phase II and a topological trivial canted AF insulator (Phase IV). The blue solid line is a second-order phase transition between Phase III and Phase IV.  Here (a-c) label three trajectories discussed in Fig. \ref{spin} and Fig. \ref{parameters}}
\label{PD}
\end{figure}

Therefore, the mean-field Hamiltonian with a collinear AF order
corresponds to a free-fermion Hamiltonian $\hat{H}_\text{H}$ with a
modified spin-dependent effective $M$, and the mean-field Hamiltonian
with a canted AF order corresponds to a free-fermion Hamiltonian with
both $M$ and $\phi$ in $\hat{H}_\text{H}$ modified. Thus, the
noninteracting phase diagram in Fig. \ref{spin}(a) is helpful for
understanding the mean-field phases, with $M$ and $\phi$ replaced by
effective parameters determined by magnetic order. There emerge two
different scenarios about how magnetic orders drive transition
between topological nontrivial and trivial insulators.

(i) If AF order increases continuously as interaction strength $U$
increases, the mean-field Hamiltonian will evolve continuously cross
the phase boundary from $\mathcal{C}=2$ to $\mathcal{C}=0$ insulator.
Inevitably, there will be a topological transition at which gapless
fermions and a finite AF order coexist.

(ii) A first-order transition occurs as $U$ increases, at which a
jump of AF order brings the system from $\mathcal{C}=2$ regime in the
phase diagram to $\mathcal{C}=0$ regime.

\textit{Mean-field Phase Diagram.} A self-consistent mean-field
calculation is conducted to determine the phase diagram. For
simplicity, we first consider the situation with $M=0$ and
$\phi=\pi/2$ in $\hat{H}_\text{H}$ in Eq. \ref{Haldane}. In our
calculation, we enlarge the unit cell to six sites of each honeycomb,
and no further assumption for order parameter ${\bf m}_i$ at these
six sites are imposed. (i.e. totally $18$ parameters are determined
from self-consistent iterations.) Enlarging the unit cell in the
magnetic ordered phase turns out to be crucial for obtaining the
state with lower energy and establishing the correct picture as
discussed below. After we obtain the self-consistent solution, we can
straightforwardly compute the single-particle gap for fermions,
scalar spin chirality order and the Chern number for mean-field
ground state \cite{supple}. The resulting phase diagram is shown in
Fig. 2, which contains both two scenarios of phase transition,
depending on the ratio $t_2/t_1$, as well as four different phases:
I. topological band insulator with no AF order ($|{\bf m}|=0$,
$\mathcal{C}=2$); II. topological AF insulator with collinear AF
order ($|{\bf m}|\neq 0$, $\mathcal{C}=2$ and $\mathcal{S}=0$); III.
trivial AF insulator with collinear AF order ($|{\bf m}|\neq 0$,
$\mathcal{C}=0$ and $\mathcal{S}=0$), and IV. trivial AF insulator
with canted AF order ($|{\bf m}|\neq 0$, $\mathcal{C}=0$ and
$\mathcal{S}\neq 0$).

\begin{figure}[t]
\includegraphics[width=3.45 in]
{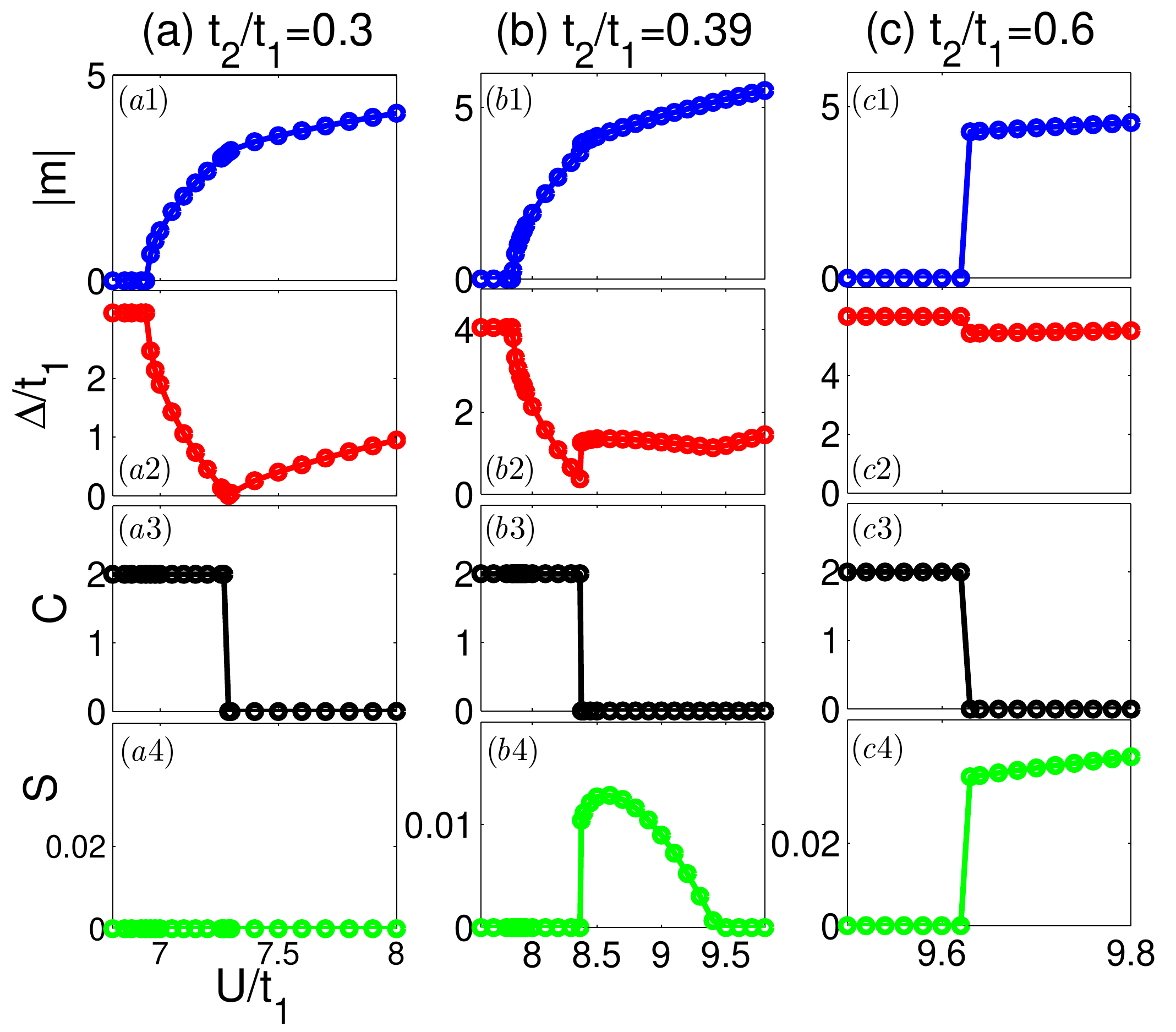} \caption{The magnetic order $|{\bf m}|$,  gap of fermion
excitation $\Delta$, Chern number $\mathcal{C}$ and scalar spin
chirality order $\mathcal{S}$ as a function of $U/t_1$ for three
different value of $t_2/t_1$, as labeled by (a-c) in Fig. \ref{PD}. }
\label{parameters}
\end{figure}

First, for small $t_2/t_1$, such as the trajectory labelled by (a) in
Fig. \ref{PD}, as $U/t_1$ increases, the system first undergoes a
second-order phase transition across which a collinear AF order
develops continuously (Fig. \ref{parameters}(a1)). As such a magnetic
order increases, the mean-field Hamiltonian $H_\text{MF}$ acquires a
$M$ ($-M$) term for spin-up (down), which suppresses the single
particle gap at $K$ ($K^\prime$) point (Fig. \ref{parameters}(a2)).
Thus, $H_\text{MF}$ undergoes a trajectory as labeled by (a) in Fig.
\ref{spin}(a). As $|{\bf m}|$ increases to a certain value, the single
particle gap closes at $K$ ($K^\prime$) point, beyond which the
mean-field ground becomes a topological trivial one (i.e.
$\mathcal{C}=0$, Fig. \ref{parameters}(a3)). Along this trajectory,
the spin chirality $\mathcal{S}$ is always zero (Fig.
\ref{parameters}(a4)).

Secondly, for intermediate $t_2/t_1$, such as the trajectory labelled
by (b) in Fig. \ref{PD}, as $U/t_1$ increases, the system first
develops a collinear AF order (Fig. \ref{parameters}(b1)). Then,
instead of reaching a topological transition, the system undergoes a
first-order transition across which the magnetic order becomes
canted. This is accompanied by a jump of spin chirality order
$\mathcal{S}$ (Fig. \ref{parameters} (b4)). This corresponds to a
discontinuous change of effective $M$ and $\phi$ in
$\hat{H}_\text{MF}$, as shown by trajectory labeled by (b) in Fig.
\ref{spin}(a). Consequently, the system jumps from a topological
phase to a topological trivial phase (Fig. \ref{parameters}(b3)), and
the gap closing point is avoided (Fig. \ref{parameters}(b2)). This
canted AF order can also be understood by a ring-exchange spin model
in term of local spin picture. As $U/t_1$ further increases, the
effective ring exchange is suppressed, $\mathcal{S}$ gradually
decreases and the system returns to a collinear AF insulator (Fig.
\ref{parameters}(b4)).

Finally, for large $t_2/t_1$, such as the trajectory labelled by (c)
in Fig. \ref{PD}, as $U/t_1$ increases, a first-order transition
directly brings the system from a topological band insulator to a
trivial canted AF insulator, across which $|{\bf m}|$, $\mathcal{C}$,
$\mathcal{S}$ all display discontinuity (Fig.
\ref{parameters}(c1),(c3),(c4)), and the fermion gap $\Delta$ remains
finite all through (Fig. \ref{parameters}(c2)).

\textit{Fluctuations Beyond Mean-field.} Here we focus on the
topological transition from Phase II to Phase III, at which the
gapless fermions coexist with gapless Goldstone spin-wave mode of AF
order. This invites the question whether the spin-wave fluctuation will
change the critical behavior. To answer this question, we introduce the
following low-energy theory with action
\begin{align}
&S=\int dt d^2{\bf r} \left( \mathcal{L}_\text{n}+\mathcal{L}_\text{f}+\mathcal{L}_\text{I}\right)\\
&\mathcal{L}_\text{n}=\frac{1}{2g}\left[(\partial_t{\bf n})^2-c^2(\nabla{\bf n})^2\right]\\
&\mathcal{L}_\text{f}= \Psi^\dag\left[i\partial_t+ iv_\text{F}\tau_z\sigma_x\partial_x + iv_\text{F}\sigma_y \partial_y - m\tau_z \sigma_z  \right]\Psi \label{fermion-lagrangian} \\
&\mathcal{L}_\text{I}= -\lambda\Psi^\dag  [\sigma_z\otimes ({\bf
n}\cdot {\bf s})]\Psi.
\end{align}
where $\mathcal{L}_\text{n}$ is a nonlinear sigma model that
describes the low-energy fluctuation of AF N\'eel order ${\bf n}$, and
the eight-component object $\Psi\equiv \Psi_{\alpha s\sigma}$
describes the Dirac fermion nearby two valleys at $K$ and $K^\prime$
points, in which $s=\uparrow$ ($\downarrow$) denotes the spin up
(down), $\alpha=1,2$ denotes the valley, and $\sigma=A(B)$ denotes
the sublattice. The Pauli matrices $s_{x,y,z}$, $\sigma_{x,y,z}$, and
$\tau_{x,y,z}$ refer to the spin, sublattice, and valley degrees of
freedom, respectively. The parameters $c$ and $v_\text{F}$ are the
spin-wave and fermion velocities, respectively. Parameters $g$ and $\lambda$ are
coupling constants of spin fluctuation and coupling between spin and
fermions, respectively. Here $c$, $v_\text{F}$, $g$, $m$ and $\lambda$ can all
be given by microscopic parameters. In particular, in Eq.\ref{fermion-lagrangian}
the mass $m$ is given by $m=3\sqrt{3}t_2$.

\begin{figure}[t]
\includegraphics[width=3.2 in]
{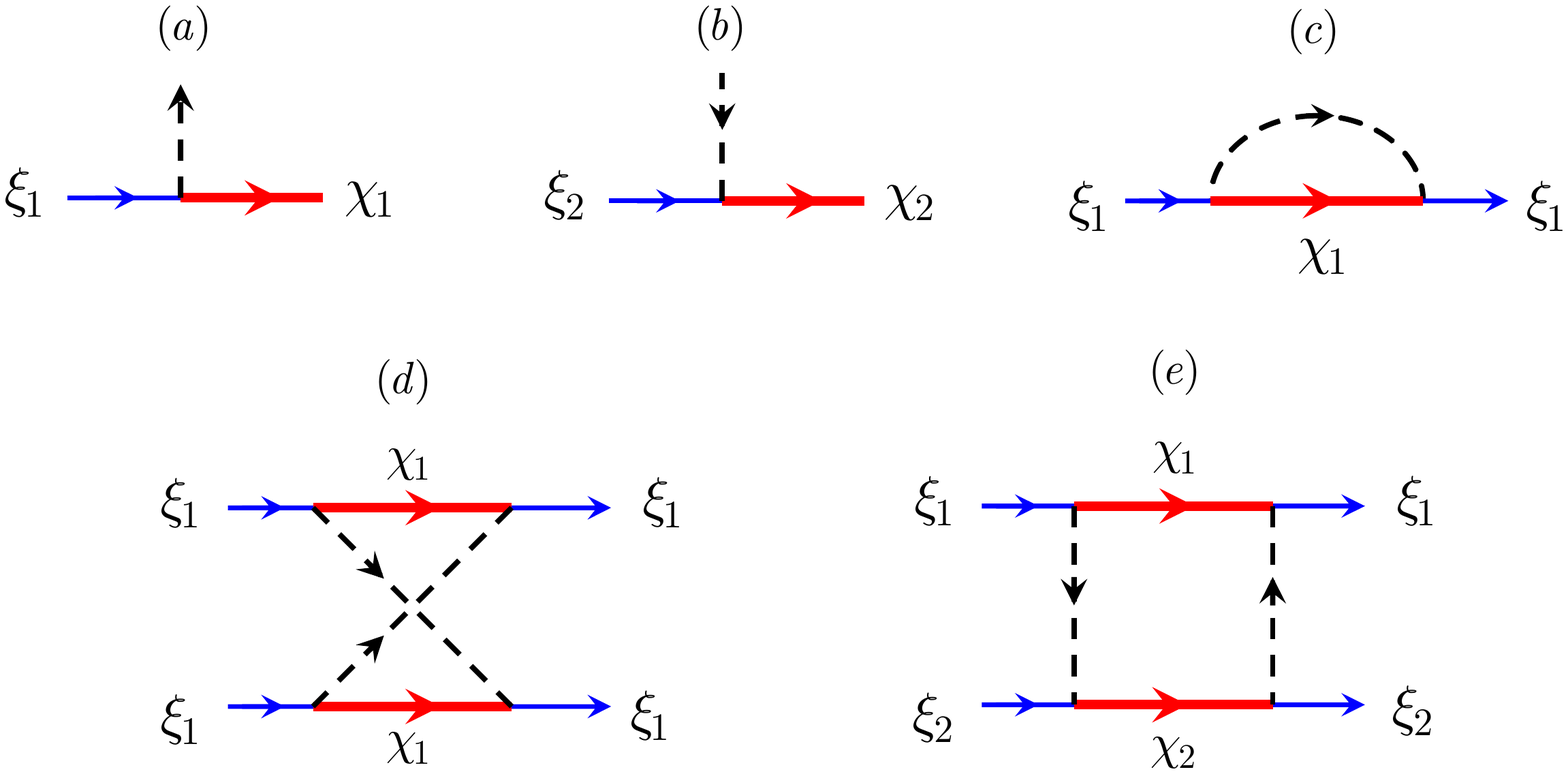} \caption{ The spin-wave-fermion vertices are shown in (a,b), and the self-energy correction for the low energy fermions is shown in (c).  The induced interaction among gapless fermions is shown in (d,e). The thin solid lines are the gapless fermion
$\xi_{\kappa}$ and the thick solid lines are the gapped fermions
$\chi_{\kappa}$. The dashed line represents spin-wave $\varphi\equiv n_x+in_y$. }
\label{diagram}
\end{figure}

With collinear AF order, we can assume that $\langle {\bf n}\rangle$
is ordered along $\hat{z}$ direction, and then take $n_z=1$ and
expand the action to the linear order of $n_x$ and $n_y$,
$\varphi=n_x+in_y$ representing a complex gapless boson field. To
bring $\mathcal{L}$ into a more convenient form, we implement a local
spin rotation $\Psi({\bf x},t)\rightarrow U({\bf x},t)\Psi({\bf
x},t)$, with $U =\exp[i(n_y s_x -n_x s_y)/2 +\cdots ]$, such that
$U^\dag ({\bf n}\cdot{\bf s})U=s_z$. Since this rotation involves
only fermions, the $\mathcal{L}_\text{n}$ term is unchanged, while
$\mathcal{L}_\text{f}+\mathcal{L}_\text{I}$ becomes
\begin{eqnarray}
 \mathcal{L}_\text{f}+\mathcal{L}_\text{I}  =   \Psi^\dag\left[iD_t+ iv_\text{F}\tau_z\sigma_x D_x +
iv_\text{F}\sigma_y D_y   \right. \nonumber \\
\left.    - m\sigma_z \tau_z  -\lambda\sigma_z s_z \right]\Psi,
\end{eqnarray} where the covariant derivative $iD_\mu = i\partial_\mu
-\frac{i}{2}(s_+\partial_\mu\varphi^* - s_-\partial_\mu\varphi)$, in
which $s_\pm =\frac{1}{2}(s_x\pm is_y)$ and $\mu=t,x,y$.  Thus, the mass term
becomes $ -\Psi^\dag \sigma_z\otimes (m\tau_z \otimes I +\lambda
I\otimes s_z)\Psi$, where $I$ denotes $2\times 2$ identity matrix. It
is therefore clear that when $\lambda=m$, namely, at the mean-field
critical point, $\Psi_{1\downarrow}$ and $\Psi_{2\uparrow}$ become
gapless (the sublattice index $\sigma$ is suppressed). Hereafter we
shall define $\xi_1\equiv \Psi_{1\downarrow}$ and
$\xi_{2}\equiv\Psi_{2\uparrow}$, which are gapless fermions, and
 $\chi_1\equiv \Psi_{1\uparrow}$ and
$\chi_{2}\equiv\Psi_{2\downarrow}$, which are gapped fermions.

The low-energy spin waves have small momentum, therefore the
spin-wave-fermion interaction given by $D_\mu$ terms does not change
the valley index, i.e. by interacting with spin waves, $\xi_{\kappa}$
can only turn into $\chi_{\kappa}$ with same $\kappa$ (Fig.
\ref{diagram}(a,b)). After integrating out the spin-wave and the
gapped fermions $\chi_\kappa$, we can obtain a self-energy correction
for low-energy fermions $\xi_\kappa$, as well as effective
interactions among the massless fermions \cite{supple}, with
corresponding diagrams shown in Fig. \ref{diagram}(d-e). To the
lowest order the self-energy takes the form of $ \Sigma
\tau_z\sigma_z$. Thus, it merely shifts the phase boundary. The
induced interaction reads
\begin{eqnarray}
\hat{V}= \int \frac{d^3 k_1}{(2\pi)^3} \frac{d^3 k_2}{(2\pi)^3}
\frac{d^3 q}{(2\pi)^3} V_q \left[ \sum_{\kappa=1,2}(\xi^\dag_{\kappa,
k_1-q} \sigma_z \xi_{\kappa, k_1 }) \right. \nonumber \\
\left. \times
(\xi^\dag_{\kappa, k_2+q }\sigma_z \xi_{\kappa,k_2}) - 2
(\xi^\dag_{1, k_1-q} \sigma_z \xi_{1, k_1 })  (\xi^\dag_{2, k_2+q
}\sigma_z \xi_{2,k_2})\right]
\end{eqnarray}
where $q=(\omega,{\bf q})$, $d^3 q$ is a shorthand notation for $\int
d{\bf q}d\omega$, and similarly for $d^3k$. Neglecting the $q$
dependence of $V(q)$, we have $V(q)= - u\equiv - (1/2)^4 g^2
c\Lambda^3/ 6\pi^2 m^2$ under certain approximation\cite{supple},
where $\Lambda$ is a momentum cutoff.

This spin-wave-induced interaction, if sufficiently strong, can open
up a gap at the nominal critical point. To see this fact, we only
need to do a mean-field approximation of $\hat{V}$. We find that if
$u>u_c\equiv \pi v_F/\Lambda$, the gapless ``ground-state'' at
$\lambda=m$ is unstable towards the dynamical generation\cite{nambu}
of a mass term $\pm\Delta\xi^\dag\sigma_z s_z\xi$, with $\Delta=  \pi
v_F^2(1/u_c- 1/u)$. Away from the $\lambda=m$ point, the $-$ sign
($+$ sign) in $\pm\Delta\xi^\dag\sigma_z s_z\xi$ is selected at
$\lambda=m+0^+$ ($\lambda=m- 0^+$). That is to say, the generated
mass jumps by $2\Delta$ across the mean-field transition point
$\lambda=m$, thus, the nominal gap closing of fermions is avoided,
and the transition becomes a first-order one.

Finally, we remark that this physics triggered by massless spin-wave
has no counterpart in the Kane-Mele-Hubbard model, because the
$SU(2)$ spin rotational symmetry is explicitly broken there, thus the
Goldstone mode is absent therein.

\textit{Final remarks.} Recent cold atom realization of the Haldane
model can be naturally described by this HH model. In fact, in the
experiment reported in Ref. \cite{Shaking}, Mott insulator with
suppressed double-occupancy sites has been observed. Our theoretical
predications can be directly verified in this setup. In this
realization, since the most crucial next nearest hopping term is
generated by periodically shaking optical lattices, the periodic
driving will also modify the interaction term, in the order of
$1/\omega$ ($\omega$ is shaking frequency). The thermal fluctuation
of magnetic order may also be important. These effects will be left
for future investigations.

We would like to thank Yi-Zhuang You, Hong Yao, Fa Wang and Shou-Cheng Zhang for discussions. This work is supported by Tsinghua University Initiative Scientific Research Program, NSFC under Grant No.~11304175 (ZW), No.~11325418 (HZ), No.~11174176 (HZ), and NKBRSFC under Grant No. 2011CB921500 (HZ).

Note added: Upon finishing this work, we became aware of Ref.\cite{hickey}, in which the same model is studied.

\begin{widetext}
\section{ Supplemental material }

\subsection{Mean-field theory}

The Hamiltonian of the Haldane-Hubbard model is given by%
\begin{eqnarray}
\hat{H} &=&-t_{1}\sum_{\left\langle ij\right\rangle ,s}\left( \hat{c}%
_{i,s}^{\dag }\hat{c}_{j,s}+\text{h.c.}\right) -t_{2}\sum_{\left\langle
\left\langle ij\right\rangle \right\rangle ,s}\left( e^{i\phi _{ij}}\hat{c}%
_{i,s}^{\dag }\hat{c}_{j,s}+\text{h.c.}\right)   \notag \\
&&-M\sum_{i,s}\epsilon _{i}\hat{c}_{i,s}^{\dag }\hat{c}_{i,s}+U\sum_{i}\hat{n%
}_{i,\uparrow }\hat{n}_{i,\downarrow }.
\end{eqnarray}%
In the following we will only consider the $M=0$ case. At the
mean-field level, we decompose the on-site interaction term as
\begin{eqnarray}
U\sum_{i}\hat{n}_{i,\uparrow }\hat{n}_{i,\downarrow } &=&\frac{1}{2}U\hat{N}-%
\frac{2}{3}\sum_{i}\mathbf{S}_{i}^{2} \\
&\approx &\frac{1}{2}U\hat{N}+\sum_{i}\left( -\mathbf{m}_{i}\cdot \mathbf{S}%
_{i}+\frac{3\mathbf{m}_{i}^{2}}{8U}\right) ,
\end{eqnarray}%
where $\mathbf{S}_{i}=\frac{1}{2}\sum\limits_{ss^{\prime }}\hat{c}%
_{i,s}^{\dag }\mathbf{\sigma }_{ss^{\prime }}\hat{c}_{i,s^{\prime }}$ is the
spin operator and $\mathbf{m}_{i}=4U\left\langle \mathbf{S}_{i}\right\rangle
/3$ is the on-site magnetic order parameter. The mean-field Hamiltonian can
be constructed as
\begin{eqnarray}
\hat{H}_{\mathrm{MF}} &=&-t_{1}\sum_{\left\langle ij\right\rangle ,s}\left(
\hat{c}_{i,s}^{\dag }\hat{c}_{j,s}+\text{h.c.}\right)
-t_{2}\sum_{\left\langle \left\langle ij\right\rangle \right\rangle
,s}\left( e^{i\phi _{ij}}\hat{c}_{i,s}^{\dag }\hat{c}_{j,s}+\text{h.c.}%
\right)   \notag \\
&&-\sum_{i}\left\{ m_{i}^{z}\left( \hat{c}_{i,\uparrow }^{\dag }\hat{c}%
_{i,\uparrow }-\hat{c}_{i,\downarrow }^{\dag }\hat{c}_{i,\downarrow }\right)
+m_{i}^{x}\left( \hat{c}_{i,\uparrow }^{\dag }\hat{c}_{i,\downarrow }+\hat{c}%
_{i,\downarrow }^{\dag }\hat{c}_{i,\uparrow }\right) -im_{i}^{y}\left( \hat{c%
}_{i,\uparrow }^{\dag }\hat{c}_{i,\downarrow }-\hat{c}_{i,\downarrow }^{\dag
}\hat{c}_{i,\uparrow }\right) \right\} .  \label{H_mf}
\end{eqnarray}%
This Hamiltonian is quadratic form and can be directly diagonalized.
As explained in the main text,
we enlarge the unit cell to a full hexagon containing six sites. We do a Fourier transformation%
\begin{equation*}
\hat{c}_{\alpha ,s}\left( \mathbf{k}\right) =\frac{1}{\sqrt{\mathcal{N}}}%
\sum\limits_{\mathbf{R}}e^{-i\mathbf{k\cdot R}}\hat{c}_{\alpha
,\sigma }\left( \mathbf{R}\right),
\end{equation*}%
where $\mathbf{R}$ is the position of the unit cell, $\alpha
=A1,B1,A2,B2,A3,B3$ denote the sublattices and $\mathcal{N}$ is the
total number of unit cells. The Hamiltonian can be transformed into
momentum
space:%
\begin{equation}
\hat{H}_{\mathrm{MF}}=\sum\limits_{\mathbf{k\in }\mathrm{BZ}}\Psi ^{\dag
}\left( \mathbf{k}\right) H\left( \mathbf{k}\right) \Psi \left( \mathbf{k}%
\right) ,
\end{equation}%
where $\Psi ^{\dag }\left( \mathbf{k}\right) $ is a 12-component spinor:%
\begin{equation}
\Psi ^{\dag }\left( \mathbf{k}\right) =\left( \hat{c}_{A1,\uparrow }^{\dag },%
\hat{c}_{B1,\uparrow }^{\dag },\hat{c}_{A2,\uparrow }^{\dag },\hat{c}%
_{B2,\uparrow }^{\dag },\hat{c}_{A3,\uparrow }^{\dag }\text{,}\hat{c}%
_{B3,\uparrow }^{\dag }\text{,}\hat{c}_{A1,\downarrow }^{\dag }\text{,}\hat{c%
}_{B1,\downarrow }^{\dag }\text{,}\hat{c}_{A2,\downarrow }^{\dag }\text{,}%
\hat{c}_{B2,\downarrow }^{\dag }\text{,}\hat{c}_{A3,\downarrow }^{\dag }%
\text{,}\hat{c}_{B3,\downarrow }^{\dag }\right)
\end{equation}%
Diagonalizing the matrix $H\left( \mathbf{k}\right) $, one can obtain the
energy band for the mean-field Hamiltonian%
\begin{equation}
\sum\limits_{ij}U_{\mu i}^{\dag }H_{ij}\left( \mathbf{k}\right) U_{j\nu
}=\delta _{\mu \nu }E_{\mu }\left( \mathbf{k}\right) .
\end{equation}%
Therefore the diagonalized mean-field Hamiltonian is
\begin{equation*}
\hat{H}_{\mathrm{MF}}=\sum\limits_{\mathbf{k\in }\mathrm{BZ}%
}\sum\limits_{\mu }\Phi _{\mu }^{\dag }\left( \mathbf{k}\right) E_{\mu
}\left( \mathbf{k}\right) \Phi _{\mu }\left( \mathbf{k}\right) ,
\end{equation*}%
where $\Phi _{\mu }\left( \mathbf{k}\right) =\sum\limits_{i}U_{\mu i}^{\dag
}\Psi _{i}\left( \mathbf{k}\right) $ is the fermion operator of each band.
For the half-filling case, the ground state is the full-filling of the
lowest 6 bands,%
\begin{equation}
\left\vert \mathrm{GS}\right\rangle =\prod\limits_{\mu =1}^{6}\prod\limits_{%
\mathbf{k\in }\mathrm{BZ}}\Phi _{\mu }^{\dag }\left( \mathbf{k}\right)
\left\vert 0\right\rangle .
\end{equation}%
The magnetization $\mathbf{m}_{\alpha }$ can be calculated from the ground
state as%
\begin{eqnarray*}
m_{\alpha }^{z} &=&\frac{1}{2\mathcal{N}}\sum\limits_{\mathbf{k\in }\mathrm{%
BZ}}\left\langle \hat{c}_{\alpha ,\uparrow }^{\dag }\left( \mathbf{k}\right)
\hat{c}_{\alpha ,\uparrow }\left( \mathbf{k}\right) -\hat{c}_{\alpha
,\downarrow }^{\dag }\left( \mathbf{k}\right) \hat{c}_{\alpha ,\downarrow
}\left( \mathbf{k}\right) \right\rangle _{\mathrm{GS}}, \\
m_{\alpha }^{x} &=&\frac{1}{2\mathcal{N}}\sum\limits_{\mathbf{k\in }\mathrm{%
BZ}}\left\langle \hat{c}_{\alpha ,\uparrow }^{\dag }\left( \mathbf{k}\right)
\hat{c}_{\tau ,\downarrow }\left( \mathbf{k}\right) -\hat{c}_{\alpha
,\downarrow }^{\dag }\left( \mathbf{k}\right) \hat{c}_{\alpha ,\uparrow
}\left( \mathbf{k}\right) \right\rangle _{\mathrm{GS}}, \\
m_{\alpha }^{y} &=&\frac{1}{2i\mathcal{N}}\sum\limits_{\mathbf{k\in }\mathrm{%
BZ}}\left\langle \hat{c}_{\alpha ,\uparrow }^{\dag }\left( \mathbf{k}\right)
\hat{c}_{\tau ,\downarrow }\left( \mathbf{k}\right) +\hat{c}_{\alpha
,\downarrow }^{\dag }\left( \mathbf{k}\right) \hat{c}_{\alpha ,\uparrow
}\left( \mathbf{k}\right) \right\rangle _{\mathrm{GS}}.
\end{eqnarray*}%
Since there are six sites in each unit cell, one has to
self-consistently calculate 18 parameters of $\mathbf{m}_{\alpha }$.
We numerically calculate the ground state and iterate until the
magnetization at each site converges. We then calculate the spin
chirality order defined by
\begin{equation*}
\mathcal{S}=\left\langle \mathbf{\hat{S}}_{i}\right\rangle \cdot
\left(
\left\langle \mathbf{\hat{S}}_{j}\right\rangle \times \left\langle \mathbf{\hat{S}}%
_{k}\right\rangle \right) ,
\end{equation*}%
The Chern number of interacting quantum anomalous Hall insulators has
been defined in Ref.\cite{simplified}. At the mean-field level it is
reduced to the Chern number of the mean-field wavefunction.  We
numerically calculate this Chern number using the algorithm given in
Ref. \cite{cn}.

\subsection{Effective Field Theory}

Near the phase boundary between Phase II to Phase III, we introduce
the
following low-energy thoery with action%
\begin{eqnarray*}
S &=&\int dtd^{2}\mathbf{r}\left( \mathcal{L}_{\mathrm{n}}+\mathcal{L}_{%
\mathrm{f}}+\mathcal{L}_{\mathrm{I}}\right) , \\
\mathcal{L}_{\mathrm{n}} &=&\frac{1}{2g}\left[ \left( \partial _{t}\mathbf{n}%
\right) ^{2}-c^{2}\left( \nabla \mathbf{n}\right) ^{2}\right] , \\
\mathcal{L}_{\mathrm{f}} &=&\Psi ^{\dag }\left[ i\partial _{t}+v_{F}\tau
_{z}\sigma _{x}i\partial _{x}+v_{F}\sigma _{y}i\partial _{y}-m\tau
_{z}\sigma _{z}\right] \Psi , \\
\mathcal{L}_{\mathrm{I}} &=&-\lambda \Psi ^{\dag }\left[ \sigma _{z}\otimes
\left( \mathbf{n\cdot s}\right) \right] \Psi .
\end{eqnarray*}%
The parameters are explained in the text. With the collinear AF
order, we can assume that the $\mathbf{n}$ is ordered along
$\hat{z}$, direction, namely, $\vec{n}\approx (0, 0, 1)$. We expand
$\mathcal{L}_{n}$ to the linear
order of $n_{x}$ and $n_{y}$ as%
\begin{equation}
\mathcal{L}_{\mathrm{n}}=\frac{1}{2g}\sum\limits_{i=x,y}\left[ \left(
\partial _{t}n_{i}\right) ^{2}-c^{2}\left( \nabla n_{i}\right) ^{2}\right] .
\end{equation}%
Then we make a spin rotation, $\Psi (x)\rightarrow U(x)\Psi (x)$, with $%
U(x)=\exp \left[ i\left( n_{y}(x)s_{x}-n_{x}(x)s_{y}\right)/2 +\cdots \right] $%
, such that%
\begin{equation*}
U^{\dag }(x)\left( \mathbf{n\cdot s}\right) U(x)=s_{z}.
\end{equation*}%
Then the last term of the Lagrangian becomes
\begin{equation}
\mathcal{L}_{\mathrm{I}}=-\lambda \Psi ^{\dag }\left( \sigma _{z}\otimes
s_{z}\right) \Psi .
\end{equation}%
It is a mass term of the fermions, so it could be absorbed into the
action of fermions. We have $\mathcal{L}_{\mathrm{f}} +
\mathcal{L}_{\mathrm{I}}$
\begin{equation}
\mathcal{L}_{\mathrm{f}}=\Psi ^{\dag }\left[ iD_{t}+v_{F}\tau _{z}\sigma
_{x}iD_{x}+v_{F}\sigma _{y}iD_{y}-m\tau _{z}\sigma _{z}-\lambda \sigma
_{z}s_{z}\right] \Psi ,
\end{equation}%
where the covariant derivative is given by:%
\begin{eqnarray*}
D_{\mu } &=&\partial _{\mu }+U^{\dag }\partial _{\mu }U \\
&=&\partial _{\mu }+\frac{1}{2}\left( s_{-}\partial _{\mu }\varphi
-s_{+}\partial _{\mu }\varphi ^{\ast }\right) ,
\end{eqnarray*}%
where $s_{\pm }=\frac{1}{2}\left( s_{x}\pm is_{y}\right) $, and
$\varphi =n_{x}+in_{y}$. This $\mathcal{L}_{\mathrm{f}} +
\mathcal{L}_{\mathrm{I}}$ can be written explicitly as

\begin{eqnarray}
\mathcal{L}_{\mathrm{f}} &=&\Psi _{1\uparrow }^{\dag }\left[ i\partial
_{t}+v_{F}\sigma _{x}i\partial _{x}+v_{F}\sigma _{y}i\partial _{y}-\left(
m+\lambda \right) \sigma _{z}\right] \Psi _{1\uparrow }+\Psi _{1\downarrow
}^{\dag }\left[ i\partial _{t}+v_{F}\sigma _{x}i\partial _{x}+v_{F}\sigma
_{y}i\partial _{y}-\left( m-\lambda \right) \sigma _{z}\right] \Psi
_{1\downarrow }  \notag \\
&&+\Psi _{2\uparrow }^{\dag }\left[ i\partial _{t}-v_{F}\sigma _{x}i\partial
_{x}+v_{F}\sigma _{y}i\partial _{y}+\left( m-\lambda \right) \sigma _{z}%
\right] \Psi _{2\uparrow }+\Psi _{2\downarrow }^{\dag }\left[ i\partial
_{t}-v_{F}\sigma _{x}i\partial _{x}+v_{F}\sigma _{y}i\partial _{y}+\left(
m+\lambda \right) \sigma _{z}\right] \Psi _{2\downarrow }  \notag \\
&&+\frac{1}{2}\Psi _{1\downarrow }^{\dag }\left( i\partial _{t}\varphi
+v_{F}\sigma _{x}i\partial _{x}\varphi +v_{F}\sigma _{y}i\partial
_{y}\varphi \right) \Psi _{1\uparrow }-\frac{1}{2}\Psi _{1\uparrow }^{\dag
}\left( i\partial _{t}\varphi ^{\ast }+v_{F}\sigma _{x}i\partial _{x}\varphi
^{\ast }+v_{F}\sigma _{y}i\partial _{y}\varphi ^{\ast }\right) \Psi
_{1\downarrow }  \notag \\
&&+\frac{1}{2}\Psi _{2\downarrow }^{\dag }\left( i\partial _{t}\varphi
-v_{F}\sigma _{x}i\partial _{x}\varphi +v_{F}\sigma _{y}i\partial
_{y}\varphi \right) \Psi _{2\uparrow }-\frac{1}{2}\Psi _{2\uparrow }^{\dag
}\left( i\partial _{t}\varphi ^{\ast }-v_{F}\sigma _{x}i\partial _{x}\varphi
^{\ast }+v_{F}\sigma _{y}i\partial _{y}\varphi ^{\ast }\right) \Psi
_{2\downarrow }.
\end{eqnarray}

When $\lambda =m$, namely, at the mean field critical point, $\mathcal{L}_{%
\mathrm{f}}$ becomes%
\begin{eqnarray}
\mathcal{L}_{\mathrm{f}} &=&\Psi _{1\uparrow }^{\dag }\left[ i\partial
_{t}+v_{F}\sigma _{x}i\partial _{x}+v_{F}\sigma _{y}i\partial _{y}-2m\sigma
_{z}\right] \Psi _{1\uparrow }+\Psi _{1\downarrow }^{\dag }\left[ i\partial
_{t}+v_{F}\sigma _{x}i\partial _{x}+v_{F}\sigma _{y}i\partial _{y}\right]
\Psi _{1\downarrow }  \notag \\
&&+\Psi _{2\uparrow }^{\dag }\left[ i\partial _{t}-v_{F}\sigma _{x}i\partial
_{x}+v_{F}\sigma _{y}i\partial _{y}\right] \Psi _{2\uparrow }+\Psi
_{2\downarrow }^{\dag }\left[ i\partial _{t}-v_{F}\sigma _{x}i\partial
_{x}+v_{F}\sigma _{y}i\partial _{y}+2m\sigma _{z}\right] \Psi _{2\downarrow }
\notag \\
&&+\frac{1}{2}\Psi _{1\downarrow }^{\dag }s_{-}\left( i\partial _{t}\varphi
+v_{F}\sigma _{x}i\partial _{x}\varphi +v_{F}\sigma _{y}i\partial
_{y}\varphi \right) \Psi _{1\uparrow }-\frac{1}{2}\Psi _{1\uparrow }^{\dag
}s_{+}\left( i\partial _{t}\varphi ^{\ast }+v_{F}\sigma _{x}i\partial
_{x}\varphi ^{\ast }+v_{F}\sigma _{y}i\partial _{y}\varphi ^{\ast }\right)
\Psi _{1\downarrow }  \notag \\
&&+\frac{1}{2}\Psi _{2\downarrow }^{\dag }s_{-}\left( i\partial _{t}\varphi
-v_{F}\sigma _{x}i\partial _{x}\varphi +v_{F}\sigma _{y}i\partial
_{y}\varphi \right) \Psi _{2\uparrow }-\frac{1}{2}\Psi _{2\uparrow }^{\dag
}s_{+}\left( i\partial _{t}\varphi ^{\ast }-v_{F}\sigma _{x}i\partial
_{x}\varphi ^{\ast }+v_{F}\sigma _{y}i\partial _{y}\varphi ^{\ast }\right)
\Psi _{2\downarrow }.
\end{eqnarray}%
We can see that $\Psi _{1\downarrow }$ and $\Psi _{2\uparrow }$
become gapless, while $\Psi _{1\uparrow }$ and $\Psi _{2\downarrow }$
are gapped. We redefine
\begin{eqnarray*}
\xi _{1} &=&\Psi _{1\downarrow },\text{ }\xi _{2}=\Psi _{2\uparrow },\text{ }%
\chi _{1}=\Psi _{1\uparrow },\text{ }\chi _{2}=\Psi _{2\downarrow } \\
\bar{\xi}_{1} &=&\Psi _{1\downarrow }^{\dag }\sigma _{z},\text{ }\bar{\xi}%
_{2}=\Psi _{2\uparrow }^{\dag }\sigma _{z},\text{ }\bar{\chi}_{1}=\Psi
_{1\uparrow }^{\dag }\sigma _{z},\text{ }\bar{\chi}_{2}=\Psi _{2\downarrow
}^{\dag }\sigma _{z} \\
\varphi  &=&n_{x}+in_{y},\text{ }\varphi ^{\ast }=n_{x}-in_{y},
\end{eqnarray*}%
then the total action becomes%
\begin{eqnarray*}
\mathcal{L}_{\mathrm{n}} &=&\frac{1}{2g}\left( \left\vert \partial
_{t}\varphi \right\vert ^{2}-c^{2}\left\vert \nabla \varphi \right\vert
^{2}\right)  \\
\mathcal{L}_{\mathrm{f}} &=&\bar{\xi}_{1}\left( \gamma _{0}i\partial
_{t}+v_{F}\gamma _{1}i\partial _{x}+v_{F}\gamma _{2}i\partial _{y}\right)
\xi _{1}+\bar{\xi}_{2}\left( \gamma _{0}i\partial _{t}-v_{F}\gamma
_{1}i\partial _{x}+v_{F}\gamma _{2}i\partial _{y}\right) \xi _{2} \\
&&+\bar{\chi}_{1}\left( \gamma _{0}i\partial _{t}+v_{F}\gamma _{1}i\partial
_{x}+v_{F}\gamma _{2}i\partial _{y}-2m\right) \chi _{1}+\bar{\chi}_{2}\left(
\gamma _{0}i\partial _{t}-v_{F}\gamma _{1}i\partial _{x}+v_{F}\gamma
_{2}i\partial _{y}+2m\right) \chi _{2}, \\
\mathcal{L}_{\mathrm{I}} &=&\frac{1}{2}\bar{\xi}_{1}\left( \gamma
_{0}i\partial _{t}\varphi +v_{F}\gamma _{1}i\partial _{x}\varphi
+v_{F}\gamma _{2}i\partial _{y}\varphi \right) \chi _{1}-\frac{1}{2}\bar{\chi%
}_{1}\left( \gamma _{0}i\partial _{t}\varphi ^{\ast }+v_{F}\gamma
_{1}i\partial _{x}\varphi ^{\ast }+v_{F}\gamma _{2}i\partial _{y}\varphi
^{\ast }\right) \xi _{1} \\
&&+\frac{1}{2}\bar{\chi}_{2}\left( \gamma _{0}i\partial _{t}\varphi
-v_{F}\gamma _{1}i\partial _{x}\varphi +v_{F}\gamma _{2}i\partial
_{y}\varphi \right) \xi _{2}-\frac{1}{2}\bar{\xi}_{2}\left( \gamma
_{0}i\partial _{t}\varphi ^{\ast }-v_{F}\gamma _{1}i\partial _{x}\varphi
^{\ast }+v_{F}\gamma _{2}i\partial _{y}\varphi ^{\ast }\right) \chi _{2}.
\end{eqnarray*}%
Note that we have mapped the low energy theory into four Dirac fields
interacting with a complex scalar field. In the following, we will
integrate out the massive Dirac fields and the complex scaler field
to obtain an effective theory for the massless fermions.

The partition function is given by%
\begin{equation}
Z=\int D\xi D\chi D\varphi e^{iS_{\xi }+iS_{\chi }+iS_{\varphi }+iS_{\mathrm{%
I}}}.
\end{equation}%
Integrating out the $\chi $ and $\varphi $ field gives rise to an
effective action
for $\xi $ field%
\begin{equation}
e^{iS\mathrm{eff}}=e^{iS_{\xi }}\int D\chi D\varphi e^{iS_{\chi
}+iS_{\varphi }+iS_{\mathrm{I}}}=e^{iS_{\xi }}\left\langle e^{iS_{\mathrm{I}%
}}\right\rangle _{0},
\end{equation}%
Here $\left\langle e^{iS_{\mathrm{I}}}\right\rangle _{0}$ is the average
over free $\chi $ and $\varphi $ field, which reads%
\begin{eqnarray}
\left\langle e^{iS_{\mathrm{I}}}\right\rangle _{0} &=&1-\frac{1}{2!}%
\left\langle S_{\mathrm{I}}^{2}\right\rangle +\frac{1}{4!}\left\langle S_{%
\mathrm{I}}^{4}\right\rangle +\cdots   \notag \\
&=&\exp \left( -\frac{1}{2!}\left\langle S_{\mathrm{I}}^{2}\right\rangle +%
\frac{1}{4!}\left( \left\langle S_{\mathrm{I}}^{4}\right\rangle
-3\left\langle S_{\mathrm{I}}^{2}\right\rangle ^{2}\right) +\cdots \right)
\end{eqnarray}%
The effective action has the form of
\begin{equation}
S_{\mathrm{eff}}=S_{\xi }+\frac{i}{2!}\left\langle S_{\mathrm{I}%
}^{2}\right\rangle -\frac{i}{4!}\left( \left\langle S_{\mathrm{I}%
}^{4}\right\rangle -3\left\langle S_{\mathrm{I}}^{2}\right\rangle
^{2}\right) .  \label{S_eff}
\end{equation}

The second term in Eq. \ref{S_eff} generates a self-energy for the
gapless
fermion, which is illustrated in fig.4 (c).%
\begin{equation*}
\frac{i}{2!}\left\langle S_{\mathrm{I}}^{2}\right\rangle =\int \frac{d^{3}p}{%
\left( 2\pi \right) ^{3}}\bar{\xi}_{\kappa }\left( p\right) \Sigma _{\kappa
}\left( p\right) \xi _{\kappa }\left( p\right) .
\end{equation*}%
This self-energy has the form of
\begin{eqnarray}
\Sigma _{1,2}\left( p\right)  &=&i\int \frac{d^{3}k}{\left( 2\pi \right) ^{3}%
}\frac{\gamma _{0}k_{0}-v_{F}\gamma _{1}k_{1}-v_{F}\gamma _{2}k_{2}}{2}%
iD\left( k\right) \frac{\gamma _{0}k_{0}-v_{F}\gamma _{1}k_{1}-v_{F}\gamma
_{2}k_{2}}{2}iK_{1,2}(p-k), \\
&=&\frac{i}{4}\int \frac{d^{3}k}{\left( 2\pi \right) ^{3}}\left(
k_{0}^{2}-v_{F}^{2}k_{1}^{2}-v_{F}^{2}k_{2}^{2}\right) D\left( k\right)
K_{1,2}(p-k),
\end{eqnarray}%
where $D\left( k\right) $ is the propagator of the complex scaler field, and
$K_{1,2}(k)$ is the propagator of the massive Dirac fermions: %
\begin{eqnarray}
D\left( k\right)  &=&\frac{2g}{k_{0}^{2}-c^{2}k_{1}^{2}-c^{2}k_{2}^{2}+i%
\varepsilon } \\
K_{1,2}\left( k\right)  &=&\frac{1}{\gamma _{0}k_{0}\mp v_{F}\gamma
_{1}k_{1}-v_{F}\gamma _{2}k_{2}\mp 2m+i\varepsilon }.
\end{eqnarray}%
For low energy processes, we approximate $K_{1,2}\left( k\right)
\approx \mp \frac{1}{2m}$, so that the self energy is given by:%
\begin{equation}
\Sigma _{1,2}\left( p\right) =\mp \frac{ig}{4m}\int \frac{d^{3}k}{\left(
2\pi \right) ^{3}}\frac{k_{0}^{2}-v_{F}^{2}k_{1}^{2}-v_{F}^{2}k_{2}^{2}}{%
k_{0}^{2}-c^{2}k_{1}^{2}-c^{2}k_{2}^{2}+i\varepsilon },
\end{equation}%
One can see that it merely shifts the mean-field phase boundary
without qualitatively changing its physical properties.

The third term in the effective action \ref{S_eff}, as illustrated in
fig.4 (d)
and (e), generates an effective interaction between the gapless fermions: %
\begin{equation*}
-\frac{i}{4!}\left( \left\langle S_{\mathrm{I}}^{4}\right\rangle
-3\left\langle S_{\mathrm{I}}^{2}\right\rangle ^{2}\right) =-\frac{1}{2}\int
\frac{d^{3}q}{\left( 2\pi \right) ^{3}}\frac{d^{3}k_{1}}{\left( 2\pi \right)
^{3}}\frac{d^{3}k_{2}}{\left( 2\pi \right) ^{3}}V_{q}\left(
\sum\limits_{\kappa =1,2}\bar{\xi}_{\kappa ,k_{1}-q}\bar{\xi}_{\kappa
,k_{2}+q}\xi _{\kappa ,k_{2}}\xi _{\kappa ,k_{1}}-2\bar{\xi}_{1,k_{1}-q}\bar{%
\xi}_{2,k_{2}+q}\xi _{2,k_{2}}\xi _{1,k_{1}}\right) ,
\end{equation*}%
where the $V_{q}$ is given by:%
\begin{eqnarray*}
V_{q} &=&\frac{i}{2^{4}}\frac{1}{4m^{2}}\int \frac{d^{3}k}{\left( 2\pi
\right) ^{3}} \\
&&\times \left(
k_{0}^{2}-v_{F}^{2}k_{1}^{2}-v_{F}^{2}k_{2}^{2}\right) D\left(
k\right) \left[ \left( q_{0}+k_{0}\right) ^{2}-v_{F}^{2}\left(
q_{1}+k_{1}\right) -v_{F}^{2}\left( q_{2}+k_{2}\right) \right]
 D\left( q+k\right) .
\end{eqnarray*}%
When $q=0$, we have:
\begin{eqnarray*}
V_{0} &=&\frac{i}{2^{4}}\frac{1}{4m^{2}}\int \frac{d^{3}k}{\left( 2\pi
\right) ^{3}}\left( k_{0}^{2}-v_{F}^{2}k_{1}^{2}-v_{F}^{2}k_{2}^{2}\right)
^{2}D^{2}\left( k\right) . \\
&=&\frac{ig^{2}}{2^{4}m^{2}}\int \frac{d^{3}k}{\left( 2\pi \right) ^{3}}%
\left( \frac{k_{0}^{2}-v_{F}^{2}k_{1}^{2}-v_{F}^{2}k_{2}^{2}}{%
k_{0}^{2}-c^{2}k_{1}^{2}-c^{2}k_{2}^{2}+i\varepsilon }\right) ^{2}. \\
&=&-\frac{g^{2}c \Lambda ^{3}}{2^{4}\pi ^{2}m^{2}}\left[ \frac{1}{6}%
-\left( 1-\frac{v_{F}^{2}}{c^{2}}\right) \frac{2}{9}+\left( 1-\frac{v_{F}^{2}%
}{c^{2}}\right) ^{2}\frac{4}{45}\right]
\end{eqnarray*}%
where $\Lambda$ is a momentum cutoff. This is the effective
interaction between the fermions.

\end{widetext}
\end{document}